\documentclass[preprint,prd, nofootinbib]{revtex4-1}
\usepackage{amsmath}
\usepackage{color}
\usepackage{graphicx}
\usepackage{epsfig}
\usepackage{subfig}
\usepackage{lipsum}
\usepackage{ctable}
\usepackage{hyperref}
\usepackage{cleveref}
\crefformat{footnote}{#2\footnotemark[#1]#3}
\usepackage{physics}
\usepackage{graphicx}
\usepackage{dcolumn}
\usepackage{bm}

\begin{document}

	\title{Inflation  of the Universe by the Non-minimal $Y(R)F^2$ Models}

	\author{ \"{O}zcan SERT}
	\email{osert@pau.edu.tr}
	\affiliation{Department of Mathematics, Faculty of Arts and Sciences,
		Pamukkale
		University,  20070   Denizli, T\"{u}rkiye}


	\date{\today}

	\begin{abstract}

		\noindent
		
		

		The cosmological solutions of the non-minimal
		 $ Y (R)F^2 $ theory which are  compatible with FRW space-time are investigated.
	 In order to avoid the isotropy  violation  of a vector field, it can be considered that the presence of a  triplet of   vector fields satisfying $SO(3)$ symmetry, or the average of randomly oriented  electromagnetic  fields over the sufficiently large volume. By considering the symmetry arguments, the inflation  driven by the non-minimally coupled  vector  fields is studied.  Then the cosmological solutions  and corresponding models are obtained with the  power law, hyperbolic and hybrid scale functions.



	\end{abstract}
	
	\pacs{Valid PACS appear here}
	\maketitle
	

	\def\ba{\begin{eqnarray}}
	\def\ea{\end{eqnarray}}
	\def\w{\wedge}

	

\section{Introduction}

\noindent

Despite  the achievements  of Einstein's General Relativity in the solar system,  the greatest  mysteries of the universe at the large cosmological scale  remain unsolved. After the observations which are the accelerated expansion of the universe \cite{Perlmutter1999,Knop2003,Amanullah2010,Riess1998,Weinberg2013} and the flatness of  galactic  rotation curves  \cite{Overduin2004,Baer2015}, so many theories were proposed to solve the mysteries.  Although Einstein's theory with cosmological constant \cite{Weinberg1989} is consistent with the observational data,  it brings out the other problems related with the source of the cosmological constant \cite{Planck2016,Planck20162,Hildebrandt2016,Jong2015,Kuijken2015,Fenech2017,Riess2011,Riess2016,Ade2014}. Thus,
unless the dark matter particle  is observed directly, one can consider such modifications as   $f(R)$ gravity   in order to explain these observations, see \cite{Nojiri2011,Sotiriou2010,Capozziello2010} for some reviews.

Recent observations indicate the existence of magnetic fields in  the gravitational systems such  as galaxies and stars
\cite{Kim1990,Kim1991,Clarke2001,
	Giovannini2004,Widrow2002,Grasso2001}. The origin of the magnetic fields  is another mystery  of the universe.
It may be considered that  the large scale magnetic fields are seeded from the primordial magnetic fields. There are some mechanism to explain the generation and amplification of the magnetic fields  such as plasma vortical motion, inflation, cosmic strings, quark hadron phase transition and electroweak phase transition \cite{Grasso2001}. In the phase transition mechanism,
the  motion of dipole charge layers formed  on the surface of the phase transition   generates a current in the electron-positron plasma,
and hence leads to randomly magnetic field fluctuations on a thermal wavelength scale,  which mean the stochastic background magnetic fields. 
Therefore, on the scales much smaller than the Hubble radius,  the effect of  magnetic fields to the isotropy  can be ignored, and the stochastic magnetic fields become compatible with the FRW geometry. Then the energy-momentum tensor of the electromagnetic field turns out to be the energy-momentum tensor of the  radiation fluid during inflation phase.

 In the presence of electromagnetic fields, firstly, one can consider only the Maxwell Lagrangian minimally coupled to Einstein gravity which known as Einstein-Maxwell theory. However,  the minimal Lagrangian has  not any solution  to explain   inflation and the primordial magnetic fields without any couplings.  Then it can be considered to modify the minimal theory.   Moreover, one can consider to break the conformal invariance of the electromagnetic field Lagrangian by the $RF^2$ type couplings  to amplify the magnetic fields at inflation \cite{Turner1988}.   
Also  the  $RF^2$ type couplings  and their effects on gravity are studied  in \cite{Prasanna1971,Horndeski1976,Sert2011PRD}, and they were derived  from reduction of higher dimensional gravity theories \cite{Muller-Hoissen1988,Buchdahl1979,Dereli1990} and the calculation of vacuum polarization in a curved background \cite{Drummond1980}. Furthermore  it is important to note that they also have used to explain the generation of primordial magnetic fields   \cite{Turner1988,Lambiase2004,Lambiase2008,Mazzitelli1995,Campanelli2008,Bamba2008,AADS2017,SertAdak2019}. However, the amplitude of the produced seed magnetic fields  was very small for this case. Then, they were extended to the $R^nF^2$ type modifications in order to produce  sufficiently large  the seed magnetic fields which lead to the current values.
Since  the inflationary era has   very high gravitational and electromagnetic fields,    the non-minimal $Y(R)F^2$ type couplings may arise in  such extreme cases. 
Also they  
 can be reason of  the flatness of galactic  rotational curves \cite{Sert2011EPJC,Sert2011MPLA,Sert2012Plus,Sert2013MPLA} and they accept the regular black hole solutions \cite{Sert2016regular}.

Although the energy-momentum  tensor is anisotropic  in the presence of a vector field, there are interesting methods  to obtain isotropic solutions and  make the energy-momentum tensor compatible with the  FRW geometry. First   one is to introduce  a triplet of orthogonal  vector fields \cite{Bento1993,Bertolami1991,Bertolami19912}, that is, one can build models which lead to isotropic cosmology starting from a non-Abelian gauge theory.   Second is to consider a large number of vector fields in random spatial directions, and take the  spatial average of the  fields which was firstly proposed by Tolman and Ehrenfest \cite{Tolman1930}.  
The averaging  procedure  was applied to  the non-linear electromagnetic theory in literature extensively \cite{Ovgun2017,Ovgun2018,Otalora2018,Novello2004,Kruglov2015,Kruglov2016,Lorenci2002}.  In this procedure, it is assumed that  the wavelength of the  electromagnetic fields    is much smaller than the space-time curvature. Then  the spatial average of the fields  is considered on the small volumes, that is, the stochastic electromagnetic fields are used. Then the averaged energy-momentum tensor  in the stochastic background becomes  energy-momentum of a radiation fluid in the minimal case.
  In this study,  we look for  the isotropic solutions of the non-minimal $Y(R)F^2$  theory  inspired  by the symmetry arguments mentioned above. Then  we compare the obtained solutions with the previous anisotropic cases.

\section{Field Equations For The Symmetry Arguments} 
\noindent
We start with the following Lagrangian of the non-minimal $Y(R)F^2$ theory \cite{Sert2011MPLA,Sert2011EPJC,Sert2012Plus,Sert2013MPLA}
 \begin{eqnarray}\label{lag1}
  L(e^a,{\omega^a}_b, A) =  \frac{1}{2\kappa^2} R*1 -Y(R) F\wedge *F + \lambda_a  \wedge  T^a \;,
   \end{eqnarray}
where $\kappa$ is the gravitational coupling constant, $R$ is the Ricci scalar obtained from  the curvature 2-form $R_{ab} = d\omega_{ab} + \omega_{ac} \wedge {\omega^c}_b$ by the interior product $\iota_a$ such as $\iota_{b}\iota_a R^{ab}$ and $\omega_{ab}$ is the connection 1-form. In this Lagrangian,  the Hodge star operator $*$  maps $p$ form to $4-p$ form, and $*1= e^{0123}$ corresponds to the oriented volume element given by  the abbreviated notation 
$ e^{ab\cdots} = e^a\wedge e^b\wedge \cdots $ for   the orthonormal basis 1-forms  $e^a, e^b, \cdots$. Then the curved space-time metric is determined by $g=\eta_{ab} e^a\otimes e^b $,  where the flat Minkowski metric $\eta_{ab}$ has the diagonal elements $(-1,1,1,1)$. Here, the non-minimal modification  is constituted by multiplying  an arbitrary function of Ricci scalar $Y(R)$  with the Maxwell Lagrangian $F\wedge*F$. Here  $F=dA$ is the Maxwell tensor and $A$ is the  potential one form. Then the torsion two form $T^a = De^a= de^a +{\omega^a}_b\wedge e^b $  is constrained to zero by   the Lagrange multiplier $\lambda_a$ and it leads to the Levi-Civita connection.   
By taking the  variation of the Lagrangian, we obtain  the following gravitational field equation, 
\begin{eqnarray}\label{gfe0}
-\frac{1}{2 \kappa^2}  R^{bc} \wedge *e_{abc} =   Y(\iota_a F \wedge *F - F \wedge \iota_a *F)
+  Y_R F_{mn} F^{mn} *R_{a} & & \nonumber \\
+   D [ \iota^b d(Y_R F_{mn} F^{mn} )]\wedge *e_{ab}    ,
\end{eqnarray}
up to a closed form, where  $Y_R = {dY}/{dR}$.
The trace of the field equation gives the  following constraint 
\begin{eqnarray}\label{cond2}
Y_RF_{mn}F^{mn}= -\frac{1}{\kappa^2}\;.
\end{eqnarray}
 The constraint equation  eliminate  the last term  in the field equation (\ref{gfe0}) and the possible  instabilities and complexities of the higher order derivatives.
 Thus  we  have
the following field equations for the theory which have no more than  second order derivatives,   
 \begin{eqnarray}\label{gfe1}
 -\frac{1}{2 \kappa^2}  R^{bc} \wedge *e_{abc} =   Y(\iota_a F \wedge *F - F \wedge \iota_a *F)
 -\frac{1}{\kappa^2} *R_{a}      \;.
 \end{eqnarray}
We can also start with a Lagrangian which has the constraint (\ref{cond2}) by a Lagrange multiplier. By taking the variation of the Lagrangian we can obtain again the field equation (\ref{gfe1}). It is interesting to note that the constraint (\ref{cond2}) is not linearly independent from the field equation (\ref{gfe1}). When we take the exterior covariant derivative of the field equation (\ref{gfe1}), we obtain the constraint (\ref{cond2}),  see \cite{Sert2017} for details.   Thus the constraint corresponds to conservation of the energy-momentum of this model.
We consider   the following  flat  isotropic FRW metric,
\begin{equation}\label{metric}
g = - dt^2  +  a^2(t) ( dx^2 +   dy^2 + dz^2)
\end{equation}
where $a(t)$ is the scale factor of the universe. 
The presence of electromagnetic fields leads to a preferential direction and anisotropy  in the space-time geometry. 
 The following  averaging procedure which was firstly  proposed by Tolman and Ehrenfest  \cite{Tolman1930}  can be applied over the spatial volume to obtain  isotropic energy momentum tensor
\begin{eqnarray}
&& <E_i> = 0,  \hskip 1.5 cm <B_i> =0, \\ 
<E_iB_j> =0 && \hskip 1 cm
<E_iE_j>  = \frac{1}{3} E^2\eta_{ij} \hskip 1 cm <B_iB_j>  = \frac{1}{3} B^2\eta_{ij}\;.
\end{eqnarray}
 Here $E_i$ and $B_i$ electric and magnetic components of the electromagnetic field and   the  average of the  electric component $E_i$ is defined by \cite{Lorenci2002}
\begin{eqnarray}
<E_i> = \lim\limits_{V\rightarrow V_0}\frac{1}{V}
\int E_ia^3d^3x
\end{eqnarray}
where $V=\int a^3d^3x$ and $V_0$ is the  sufficiently large spatial volume satisfying these relations.

 By the symmetry arguments and the process, the Maxwell energy-momentum tensor becomes    
\begin{eqnarray}
 <F_a\wedge *F - F\wedge\iota_{a}*F> \  =  (\rho_F +p_F ) u_a*u +p_F*e_a    \;,
\end{eqnarray}
which is the energy-momentum tensor of the isotropic radiation fluid.
Here $u=u_a e^a$ and $u_a$ time-like, unit vector field. The energy density and pressure of the electromagnetic field are given as
\begin{eqnarray}
\rho_F&=& E^2 +B^2 \;, \\
p_F &=& \frac{1}{3}(E^2 +B^2)=\frac{\rho_F}{3}\;. 
\end{eqnarray} 
By  considering  the definitions $<E_1^2 +E_2^2 +E_3^2>= E^2$ \ and \ $<B_1^2 +B_2^2 +B_3^2>= B^2$, we calculate
\begin{eqnarray}\label{aver}
<F_{mn} F^{mn} >= 2(B^2-E^2)\;.
\end{eqnarray}

	Alternatively, it is interesting to note that the  symmetry arguments given in   \cite{Bento1993,Bertolami1991,Bertolami19912}  can be used to satisfy homogeneity and isotropy of the early universe. In these studies,  a vector potential 1-form  which has	a  triplet of  identical and orthogonal vector fields   is considered as $A=A^IT_I=  \phi(t) \delta^I_\mu dx^\mu T_I $.
	Here $I=1,2,3$ and  $T_I$ is the generator of the internal non-Abelian SU(2) gauge group space. Then this gauge potential leads to $E_1^2= E_2^2= E_3^2 = \frac{E^2}{3}$, and $B_1^2 =B_2^2 =B_3^2 = \frac{B^2}{3}$ as in the averaging procedure.

 Thus the gravitational field equation (\ref{gfe1}) turns out to be 
\begin{eqnarray}\label{gfe2}
-\frac{1}{2 \kappa^2}  R_{bc} \wedge *e^{abc} = \tau^a= Y (\rho_F +p_F ) u^a*u + Y p_F*e^a  -\frac{1}{\kappa^2}*R^a\;.
\end{eqnarray}
We use the definition of effective energy-momentum tensor  $\tau_a= \tau_{ab} *e^b$, and obtain the total effective energy density and pressure from $\rho= \tau_{\scalebox{0.6}{00}} $, $p = \tau_{\scalebox{0.6}{11}}=\tau_{\scalebox{0.6}{22}}=\tau_{\scalebox{0.6}{33}}$\;.
\begin{eqnarray}
\rho = Y\rho_F + \frac{3}{\kappa^2}\frac{\ddot{a}}{a}\;,
 \hskip 1 cm
p = Yp_F - \frac{1}{\kappa^2}( \frac{\ddot{a}}{a} +2\frac{\dot{a}^2}{a^2} ) \;
\end{eqnarray}
 Here $\dot{a} = \frac{d a}{d t}$ . Also we can write them in terms of the Hubble parameter
$
H= \frac{\dot{a}}{a}  
$ 
as
\begin{eqnarray}
\rho  = Y\rho_F  +\frac{3}{\kappa^2}(\dot{H} + H^2 ) \;,\hskip 1 cm 
p  = \frac{\rho}{3} - \frac{2}{\kappa^2}(\dot{H} + 2H^2 )\;.
\end{eqnarray}
Thus the modified gravitational field equation (\ref{gfe2}) gives us    the following two linearly dependent differential equations for the flat FRW metric (\ref{metric})
\begin{eqnarray}\label{difdenk}
\frac{3}{\kappa^2} \frac{\dot{a}^2}{a^2}  &=&  \frac{3}{\kappa^2} \frac{\ddot{a}}{a}  + Y(E^2 +B^2) \\
\frac{1}{\kappa^2}( 2\frac{\ddot{a}}{a} + \frac{\dot{a}^2}{a^2})  &=& \frac{1}{\kappa^2}( \frac{\ddot{a}}{a} + 2\frac{\dot{a}^2}{a^2})  -\frac{1}{3} Y(E^2 +B^2) 
\end{eqnarray}
which lead to only the following differential equation 
\begin{eqnarray}\label{difdenkofthemodel}
\frac{3}{\kappa^2}(\frac{\ddot{a}}{a} -\frac{\dot{a}^2}{a^2})  + Y(E^2 +B^2) =0 \;.
\end{eqnarray}
Here we note that we have also
  the constraint equation from (\ref{cond2})  
\begin{eqnarray} \label{cond3}
\frac{dY}{dR} = \frac{1}{2\kappa^2(E^2- B^2)}\;.
\end{eqnarray}
It is easy to check that  the constraint equation (\ref{cond3}) can  also be  obtained  by taking  the derivative of the differential equation (\ref{difdenkofthemodel}).
Also, the conservation of energy momentum tensor  can be calculated as
\begin{eqnarray}\label{consdif}
D\tau^a = D(Y(\rho_F+p_F)u^au_b)\wedge *e^b + d(Yp_F)\wedge *e^a -\frac{1}{\kappa^2}D*R^a = 0
\end{eqnarray}
 and  the following differential equation is
 obtained from  (\ref{consdif}) by using the condition  (\ref{cond3}) and the metric (\ref{metric}),
\begin{eqnarray}
\frac{\dot{R}}{2\kappa^2}\left( 1+ \frac{E^2+B^2}{E^2-B^2} \right)
 + Y(E^2+B^2)^{\scalebox{1.2}{.}} +4 Y(E^2+B^2)H=0\label{conservation2}\;
\end{eqnarray}
where $R=6(\frac{\dot{a}^2}{a^2} + \frac{\ddot{a}}{a})$.  We also note  that the conservation equation (\ref{conservation2})  is equivalent to the constraint equation (\ref{cond3}) and  it   satisfies the solution
\begin{eqnarray}\label{EB}
B = \frac{B_0}{a^2}\;, \hskip 1 cm E = \frac{E_0}{Y a^2} \;
\end{eqnarray}
where $E_0$ and $B_0$ are integration constants.
The non-zero electric fields  can also be  important
  at the beginning. Since it is considered  that the universe has the plasma phase with the free electric charges,   there exist non-zero electric fields    in the plasma.   Therefore we can   consider both cases with non-zero electric fields or magnetic fields.
 

For all these models which have   the electric and magnetic fields,  we have one differential equation (\ref{difdenkofthemodel}) and two unknown functions $Y(R(t))$ and $a(t)$. That is, each non-minimal function $Y(R)$ gives a scale factor $a(t)$, and vice versa.  Therefore,
instead of starting with a specific $Y(R)F^2$ model, we start with some known scale factors and determine the corresponding models.

\section{The  hyperbolic expansion}
 We firstly consider the following hyperbolic scale function
\begin{eqnarray}
a(t) = \sinh^k(\alpha t) 
\end{eqnarray}
where $k$ and $\alpha$ are positive constants.    Then the differential equation (\ref{difdenkofthemodel})  gives
the following solution for $E^2=0$
\begin{eqnarray}
Y(t) &=&  \frac{3k\alpha^2 \sinh^{4k-2}(\alpha t)}{\kappa^2B_0^2} \;,  \label{Yt1} \\
B(t) &=& B0 \sinh^{-2k}(\alpha t) \label{B1} \;.
\end{eqnarray}
On the other hand, we can also find the solution with $B^2 = 0$
\begin{eqnarray}
Y(t) = \frac{ \kappa^2 E_0^2  }{ 3k\alpha^2  \sinh^{4k -2}(\alpha t)} \;, \label{Yt2} \\
E(t) =  \frac{3k\alpha^2\sinh^{2k-2}(\alpha t) }{E_0\kappa^2} \label{E1} \;.
\end{eqnarray} 
We note that the equations (\ref{difdenk})-(\ref{EB})  and the solutions (\ref{Yt1})-(\ref{E1})
satisfy  the duality symmetry given  by $B\rightarrow -YE$, $Y\rightarrow \frac{1}{Y}$ and  $B_0\rightarrow -E_0$ \cite{Sert2013MPLA}. 


For these solutions, the Ricci curvature scalar is calculated as  
\begin{eqnarray}\label{R1}
R(t) = \frac{6\alpha^2k (2k-1)}{\sinh^2(\alpha t)} + 12\alpha^2k^2 \;.
\end{eqnarray}
By solving $\sinh{(\alpha t)}$ in terms of $R$ from (\ref{R1}) and substituting it  in  (\ref{Yt2})  and (\ref{Yt1}),    the non-minimal function $Y(R)$ can be obtained  as follows for the existence of   magnetic fields and electric fields, respectively 
\begin{eqnarray}
Y(R) =  \frac{Y_0}{B_0^2} (R-12\alpha^2k^2 )^{1-2k}\;, \hskip 1 cm
Y(R) = \frac{ E_0^2}{Y_0} (R-12\alpha^2k^2)^{2k-1}\;.
\end{eqnarray}
where  $ Y_0= \frac{3k\alpha^2}{\kappa^2 } \left( 6k\alpha^2(2k-1)\right)^{2k-1}$\;.
 We calculate the Hubble  parameter and  deceleration parameter  as
 \begin{eqnarray}
 H &=& \frac{\dot{a}}{a} = \alpha k \coth(\alpha t) = \alpha k \left(1+ (1+z)^{2/k} \right)^{1/2}\;,\\
q &=& -1 + \frac{d}{dt}(\frac{1}{H})   =  \frac{1}{k\cosh^2(\alpha t)}  -1 = \frac{1}{k + k(1+z)^{-2/k}   } -1 \label{q1}
\end{eqnarray}
which are equivalent   for the above  non-minimal functions or models.
In the last step, the parameters are rewritten in term of the cosmic redshift $z$ by using the relation $a=\frac{1}{1+z}$.  We see that the Hubble parameter  is infinity at $t=0$ and it decreases to the constant value $H= \alpha k$ as $t\rightarrow \infty $. 
 The deceleration parameter  decreases from $q(0) = \frac{1-k}{k}  $
to $-1$, and the phase transition occurs at $t=\frac{1}{\alpha} arccosh(\frac{1}{\sqrt{k}}) $.  It must be   $0<k<1$  to obtain the phase transition from deceleration to  acceleration. For $k>1$,  $q$ is negative   at the beginning of the universe. 
The effective energy density $\rho$ and pressure $p$ are calculated as 
\begin{eqnarray}
\rho &=&  \frac{3 \alpha^2 k^2 \cosh^2(\alpha t)}{\kappa^2 \sinh^2(\alpha t)}= \frac{3\alpha^2 k^2}{\kappa^2} (1+ (1+z)^{2/k})\;, \label{rho1} \\
p &=& - \frac{ \alpha^2 k( 
	3k\cosh^2(\alpha t) -2)}{\kappa^2 \sinh^2(\alpha t)} = -\frac{\alpha^2k}{\kappa^2}\left( 3k+ (3k-2)(1+z)^{2/k} \right) \;, \hskip 1 cm \; \label{p1}
\end{eqnarray}
and then we have the ratio
\begin{eqnarray}
 w = \frac{p}{\rho}=   \frac{2}{ 3k \cosh^2(\alpha t)  } -1= \frac{2(1+z)^{2/k}}{3k(1+ (1+z)^{2/k} )} -1\;.
\end{eqnarray}
It is interesting to note  that  the Hubble parameter, energy density and pressure are singular at $t=0$ or $a=0$  and they decrease to constant values later times.  This shows that the existence of the Big Bang singularity  at the starting point of the universe.  

We can write the pressure in terms of $\rho$ by using (\ref{rho1}) and (\ref{p1} as the equation of state
\begin{eqnarray}
p= -\rho  + f(\rho) \;, \hskip 1 cm f(\rho)= \frac{2}{3k}(\rho -\frac{3\alpha^2 k^2}{\kappa^2})\;. 
\end{eqnarray}
Under the condition $\abs{\frac{f(\rho)}{\rho}} \ll 1$, the inflationary parameters such as the  spectral index $n_s$, the tensor to scalar ratio $r$ and the running spectral index $\alpha_s$ can be expressed  as follows
\begin{eqnarray}
n_s \approx 1- 6\frac{f(\rho)}{\rho}\;, \hskip 1 cm r \approx 24\frac{f(\rho)}{\rho}\;, \hskip 1 cm \alpha_s \approx- 9\frac{f^2(\rho)}{\rho^2}\;
\end{eqnarray}
which was given in   \cite{Bamba2014}.
Then we can write 
\begin{eqnarray}\label{spect}
n_s = 1-\frac{r}{4} = 1-2\sqrt{-\alpha_s} = 1-\frac{4}{k\rho } ( \rho -\frac{3\alpha^2 k^2}{\kappa^2})
\end{eqnarray}
 for the model.
  Recent Planck data analysis \cite{Planck2016,Planck20162}   gives the constraints on the parameters as $n_s=0.968 \pm 0.006 (68\% CL)$, $r<0.11 (95\% CL)$ and $ \alpha_s = -0.003\pm 0.007 (68\% CL )$.
 When we take $n_s= 0.968$  in  (\ref{spect}), we obtain $r=0.128$ and $\alpha_s = 2.56\times 10^{-4}$
 which leads to 
 \begin{eqnarray}\label{fbolurho}
 \frac{f(\rho)}{\rho}=\frac{2}{3k\rho}(\rho-\frac{3\alpha^2 k^2}{\kappa^2}) = 5.3\times 10^{-3}\;.
 \end{eqnarray}
 From (\ref{fbolurho}) the parameters $\alpha$ and $k$ can be chosen properly to obtain a certain energy  density or magnetic field during the inflation.

\section{The power law expansion}

Secondly,  we take into account solutions with the well-known power law scale factor
\begin{eqnarray}
a(t) =a_0 t^n \label{tn}
\end{eqnarray}
where $n$ positive real constant. When we substitute the power law function (\ref{tn}) and the magnetic field $B$ (\ref{EB}) in  equation  (\ref{difdenkofthemodel}),
we obtain
\begin{eqnarray}
Y(t) =  \frac{3nt^{4n-2}}{\kappa^2B_0^2} \;,  \label{Yt12} \hskip 1 cm
B(t) = \frac{B_0}{t^{2n}}  \;
\end{eqnarray}
for  $E^2=0$,  we also obtain solutions with $B^2 = 0$ as result of the duality transformation \cite{Sert2013MPLA}
\begin{eqnarray}
Y(t) = \frac{E_0^2 \kappa^2 t^{2-4n}  }{ 3n} \;, \label{Yt3} \hskip 1cm
E(t) =  \frac{3n t^{2n-2}}{E_0\kappa^2} \;.
\end{eqnarray}

For the scale function (\ref{tn}), the Ricci scalar becomes 
\begin{eqnarray}\label{R}
R(t) = \frac{12n^2-6n}{t^2}\;.
\end{eqnarray}
By taking the inverse function of $R(t)$, we obtain
the non-minimal  functions depending on $R$ in the models without electric fields or magnetic fields, respectively 
\begin{eqnarray}
Y(R) =  \frac{3n( 12n^2 -6n  )^{2n-1} }{\kappa^2B_0^2} R^{1-2n}\;, \hskip 1 cm
Y(R) = \frac{\kappa^2 E_0^2 }{3n( 12n^2 -6n  )^{2n-1}}R^{2n-1} \;.
\end{eqnarray}
Then the cosmological parameters become   
\begin{eqnarray}
H = \frac{n}{t} = na_0^{1/n}(1+z)^{1/n}\;,
\hskip 3 cm
q = \frac{1}{n}  -1 \;.
\end{eqnarray}
  We note that we have the constant deceleration parameter for a specific $n$ value in these models.
 The effective energy density and pressure turn out to be  
\begin{eqnarray}
\rho =  \frac{3n^2}{\kappa^2 t^2}= \frac{3n^2a_0^{2/n}}{\kappa^2}(1+z)^{2/n}\;, \hskip 1 cm 
p=- \frac{n(3n-2)}{\kappa^2t^2}= -\frac{n(3n-2)a_0^{2/n}}{\kappa^2}(1+z)^{2/n} \;, \hskip 1 cm \; \label{rhop}
\end{eqnarray}
which gives the equation of state parameter
\begin{eqnarray}
w =    \frac{2}{ 3n } -1\;.
\end{eqnarray}
We also see that  the models  shows the Big Bang singularity  at $t=0$, since $H,\ p$ and $\rho$ goes to infinity  as $a\rightarrow 0$. 
From (\ref{rhop}) we see that
\begin{eqnarray}
p= -\rho + \frac{2\rho}{3n}
\end{eqnarray}
and by considering the Planck data and perfect fluid approach $\frac{f(\rho)}{\rho} = \frac{2}{3n} = 5.3\times 10^{-3}$,  gives
$n\approx 126$ at the inflation.
We note that the anisotropic solutions which obtained in \cite{AADS2017} with the mean scale factor $v=v_1t^{\alpha-1/3}$  have the same cosmological parameters (the mean Hubble and deceleration parameters)  with the isotropic scale factor $a= a_0t^n$, where $n=\alpha-1/3$. Therefore the analysis of the mean quantities in \cite{AADS2017} gives the same results   with  this isotropic case.  The both approach give approximately  the same  energy density. The total energy density can be written in terms of e-folds $N$. This  analysis    shows that the modified theory gives reasonable  inflationary parameters  during the inflation with the power law expansion. It is worth to note that for $n=2/3$ the model  reproduce the matter dominated universe with 
$Y(R) =\frac{2}{\kappa^2B_0^2}(\frac{4}{3})^{1/3} R^{-1/3}$ or $Y(R) =\frac{\kappa^2E_0^2}{2}(\frac{4}{3})^{-1/3} R^{1/3}$ and $R(t)= \frac{4}{3t^2}$\;.

\section{Hybrid Expansion Law}
 Let generalize the previous power law model considering by  the hybrid expansion law 
 \begin{eqnarray}
 a(t) = a_0 t^ne^{\alpha t}
 \end{eqnarray}
where $n$ and $\alpha$ positive constants.
Then  we obtain 
\begin{eqnarray}
B(t) = \frac{B_0}{a_0^2 t^{2n} e^{2\alpha t}}\;,
\hskip 1 cm
Y(t) = \frac{3na_0^4 t^{4n-2}e^{4\alpha t}}{B_0^2 \kappa^2  }\;,
\end{eqnarray}
\begin{eqnarray}
R(t) = \frac{12\alpha^2 t^2 +24\alpha n t +12n^2 -6 n}{t^2} \;. \label{R3}
\end{eqnarray}
 By solving $t$ from (\ref{R3}), we find the non-minimal function in term of $R$ for the model as 
 \begin{eqnarray}\label{YRhybrid}
 Y(R) =\frac{3na_0^4}{B_0^2\kappa^2} (\frac{12n\alpha +X }{R-12\alpha^2})^{4n-2}e^{\frac{4\alpha(12n\alpha +X)}{R-12\alpha^2} }
 \end{eqnarray}
where $X=\sqrt{(12n^2 -6n)R +72n\alpha^2}$. Furthermore,  we have also a dual model with an electric field, which obtained by  the duality transformation  $Y \rightarrow \frac{1}{Y}$.  We calculate the related parameters for these models
\begin{eqnarray}\label{Hq}
H= \alpha + \frac{n}{t} \;,\hskip 1 cm q= \frac{n}{(\alpha t +n )^2 } -1
\end{eqnarray}
then the energy density, pressure and $\omega$ can be expressed as
\begin{eqnarray}\label{rhop2}
\rho = \frac{3(\alpha t+n)^2}{\kappa^2 t^2} \;,\hskip 1 cm p = \frac{2n - 3(\alpha t+n)^2 }{\kappa^2 t^2}\;, \hskip  1 cm \omega = \frac{2n}{3(\alpha t +n)^2} -1 \;.
\end{eqnarray}
We note that  $\alpha=0 $ gives us the previous model with the well-known power law expansion   and  $n=0$ is the  exponential expansion.
Also, in the hybrid model the constants can be  fixed so that  the power law is more effective
 in the early universe. Thus the non-minimal $Y(R)F^2$ model with the non-minimal function $Y$ (\ref{YRhybrid}) (or dual of it with  $1/Y$) has the solution with  the hybrid expansion.
From (\ref{rhop2}) we obtain
\begin{eqnarray}
p= -\rho + \frac{2(3\alpha^2 -\kappa^2\rho)^2 }{n\kappa^2 (\sqrt{3\kappa^2\rho} -3\alpha)^2}\;.
\end{eqnarray}   
 By using the perfect fluid description of the inflationary  parameters,  we can write \begin{eqnarray}
 \frac{f(\rho)}{\rho} =  \frac{2(3\alpha^2 -\kappa^2\rho)^2 }{n\kappa^2 \rho (\sqrt{3\kappa^2\rho} -3\alpha)^2}\;.
 \end{eqnarray}
 By taking  the spectral index value $n_s$ from  the Planck observations  
 we get
 \begin{eqnarray}
 \frac{2(3\alpha^2 -\kappa^2\rho)^2 }{n\kappa^2 \rho (\sqrt{3\kappa^2\rho} -3\alpha)^2} = 5.3\times 10^{-3}
 \end{eqnarray}
 We see that it is possible to choose 
 the parameters of the model $\alpha$ and $n$ as appropriately to satisfy the cosmological indices. Then the model  is also give consistent solutions with observations. 

\section{Conclusions}

\noindent
  In this paper, we have shown that there are  a wide class of isotropic inflationary solutions  of
the non-minimal $Y(R)F^2$ gravity by   the symmetry arguments of the electromagnetic fields. Firstly, we have derived the modified field equations  under the averaging procedure which leads to isotropic energy momentum tensor. Here we note that the isotropy  can also be obtained by considering a triplet of vector fields satisfying $SO(3)$ symmetry \cite{Bento1993,Bertolami1991,Bertolami19912}. 
 Then we obtained  isotropic solutions of the model with the power law,   hyperbolic and hybrid  expansions which have electric and magnetic fields.  

We note that the studied anisotropic case in \cite{AADS2017} have the same features with the power-law model for the averaged field approach, and it   approaches to the isotropic case. 
 Furthermore,  the obtained isotropic solutions with power law, hyperbolic and hybrid expansions  have the Big Bang singularity at $t=0$. The deceleration parameter monotonically decreases from $q= \frac{1-k}{k}$ to the value $-1$ for the hyperbolic and hybrid expansion, while it is constant for the power law expansion.  
The constants in the model can be   chosen for satisfying  the  latest Planck  observations as consistent with the inflationary parameters. 
We pointed out that the non-minimal gravitational
coupling of the electromagnetic field  can be used to  describe the inflationary phase for certain coupling values of the model with the  magnetic monopole  field.

	The magnetic monopoles can be considered as topological defects and they can be source of inflation which known as  topological  inflation  proposed by  Linde \cite{Linde1994}  and Vilenkin \cite{Vilenkin1994}. When the size of the  defects is much larger than the Hubble radius, the topological inflation can occur in the center of the magnetic monopole for the model with the scalar field $\phi$
    which vanishes in the center of the topological defect and corresponds to a local maximum of the effective potential. 
  Then even if inflation phase of the universe ends in the surrounding space,  inflation of the monopole    continues without end.
On the other hand, in this study we consider the effects of the non-minimally coupled electromagnetic fields to gravity  without using any scalar field or  other exotic fields. 
However, in the  both models   once inflation begins, it can be  driven by the primordial  magnetic monopole fields.

\end{document}